\documentclass[conference,a4paper]{APSIPA2019}
\usepackage{multirow}
\usepackage{graphicx}   
\usepackage{amsmath}
\usepackage[psamsfonts]{amssymb}
\usepackage{amsxtra}
\usepackage{threeparttable}
\usepackage{subfigure}
\usepackage[colorlinks,linkcolor=black,anchorcolor=black,citecolor=black,urlcolor=black,CJKbookmarks=True,citebordercolor=green]{hyperref}
\usepackage{booktabs}
\usepackage{tabu}
\usepackage{bm}
\usepackage{fancyhdr}
\begin{document}
\title{End-to-End Emotional Speech Synthesis Using Style Tokens and Semi-Supervised Training}
\author{
\authorblockN{%
Pengfei Wu\authorrefmark{1}, Zhenhua Ling\authorrefmark{1},
Lijuan Liu\authorrefmark{2}, Yuan Jiang\authorrefmark{2}, Hongchuan Wu\authorrefmark{2} and Lirong Dai\authorrefmark{1}
}
\authorblockA{%
\authorrefmark{1}
University of Science and Technology of China, Hefei, China\\
E-mail:wpf0610@mail.ustc.edu.cn, \{zhling, lrdai\}@ustc.edu.cn}
\authorblockA{%
\authorrefmark{2}
iFLYTEK Research, iFLYTEK Co., Ltd., Hefei, P.R.China\\
E-mail:\{ljliu, yuanjiang, hcwu4\}@iflytek.com
}
}

\maketitle
\thispagestyle{fancy}

\begin{abstract}
This paper proposes an end-to-end emotional speech synthesis (ESS) method which adopts global style tokens (GSTs) for semi-supervised training.
This model is built based on the GST-Tacotron framework.
The style tokens are defined to present emotion categories.
A cross entropy loss function  between token weights and emotion labels  is designed to obtain the interpretability of style tokens utilizing the small portion of training data with emotion labels.
Emotion recognition experiments confirm that this method can achieve one-to-one correspondence between style tokens and emotion categories effectively.
Objective and subjective evaluation results show that our model outperforms the conventional Tacotron model for ESS when only 5\% of training data has emotion labels.
Its subjective performance is close to the Tacotron model trained using all emotion labels.

\end{abstract}

\section{Introduction}\label{introduction}
The text-to-speech (TTS) methods based on deep learning techiniques have developed rapidly in the past few years.
They can improve the naturalness of synthetic speech significantly compared with the traditional approaches.
Zen et.al. \cite{ZeSS13} made the first attempt of building acoustic models using deep neural networks (DNNs), which can model the complex dependencies between linguistic information and acoustic features.
With similar number of parameters, it achieved better naturalness than  hidden Markov model (HMM)-based one.
Since it was difficult to incorporate long time span contextual effects into DNNs \cite{44312},
recurrent neural network (RNN)-based methods \cite{YF2014SSbiLSTM} were then proposed to better capture temporal information.
In recent years, the audio generative models that can generate high quality natural-sounding speech \cite{vande2016wavenet} and
the end-to-end acoustic models which predicted acoustic features directly from phoneme or grapheme sequences \cite{Wang17tacotron,Shen18tacotron2} have also been proposed
and attracted the attention of many researchers.

How to generate speech with expected emotions is also an important topic in TTS \cite{HoferRC05interspeech}.
At the early stage of emotional speech synthesis (ESS) research, the rule-based methods were studied which controlled the speech parameters for emotion expression through manually-defined rules \cite{Cahn90thegeneration,MurrayA95FSE}. Although this approach can control a large number of parameters related with both voice source and vocal tract \cite{Schroder01ESSreview},
the resulting speech sounded unnatural and the emotion discriminability was not satisfactory.
The unit selection-based methods for ESS have also been studied. However, obtaining a large-scale emotional speech corpus for unit selection is usually difficult and costly.
Compared with unit selection, statistical parameter speech synthesis (SPSS) requires much smaller speech corpus and is far more flexible.
A lot of studies have been made to build acoustic models for ESS using hidden Markove models \cite{YamagishiOMK05,NoseYMK07} and neural networks \cite{AnLD17LSTM,Lee17TacotronEmotion,LT18DNNEmotion}.
In \cite{AnLD17LSTM}, emotion-dependent and emotion-independent acoustic models based on RNNs with long short-term memory (LSTM) units were investigated.
The end-to-end Tacotron model has also been applied to ESS in \cite{Lee17TacotronEmotion}.

All of the ESS methods mentioned above rely on supervised training, i.e., all training or adaptation utterances have explicit emotion labels. This condition is easy to satisfy when the datasets are built based on the performed emotions of actors. However, for spontaneous expressive speech, to annotate emotion labels for all utterances is difficult and costly. Therefore, some unsupervised approaches \cite{Skerry18tacotron,Wang18GSTs,AkuzawaIM18VAE} of expressive speech synthesis using neural networks have been proposed recently. Although these methods can learn speech variations without explicit annotations for expressiveness, the learned representations or latent embeddings were usually not fully controllable and had poor interpretability. As mentioned in \cite{Wang18GSTs}, one token may capture multiple attributes in speech.
Besides, the attributes that the learned tokens correspond to can't be determined at the training stage.

In this paper, we propose a semi-supervised training method using global style tokens (GSTs) \cite{Wang18GSTs} for ESS aiming at the condition that only a small portion of training data has emotion labels.
This model is built based on the GST-Tacotron framework.
A bank of tokens with the same number as emotion categories are defined and a cross entropy loss is introduced between the token weights and the emotion labels in order to establish a one-to-one correspondence between
tokens and emotions.
The model parameters are estimated by multi-task learning when only the emotion labels of a few training samples are available.
In our experiments, we found that  this method can achieve the interpretability of tokens and the accurate recognition of emotion labels.
When only 5\% of training data had emotion labels, our model achieved similar subjective performance with the Tacotron-based ESS model with fully supervised training.

\section{Related Work}
\subsection{End-to-end speech synthesis with style tokens}
Recently, end-to-end acoustic models with style tokens for speech synthesis have been proposed \cite{Skerry18tacotron,Wang18GSTs,AkuzawaIM18VAE}.
A Tacotron-based end-to-end speech synthesis architecture that learnt a latent embedding space of prosody was proposed in \cite{Skerry18tacotron}.
In this model, a reference encoder was defined to encode the prosody of reference speech into a fixed-length vector which contained the information not provided by the text and speaker identity.
The experimental results demonstrated that this encoder can transfer prosody between utterances in an almost speaker-independent fashion.
An unsupervised style modeling, control and transfer method was proposed in \cite{Wang18GSTs}. It extended Tacotron by adding a style token layer consisting of a bank of embeddings named global style tokens in it.
The embeddings were trained without explicit labels, and were learnt to model the expressiveness-related acoustic variations in speech.

\subsection{Neural network-based ESS}
LSTM-RNN models for ESS were studied in \cite{AnLD17LSTM}.
Two modeling approaches, emotion-dependent modeling and unified modeling with emotion codes, were designed.
Experimental results showed that both approaches achieved significant better naturalness of synthetic speech than HMM-based emotion-dependent modeling.
Emotion control can be achieved by interpolating or extrapolating the emotion codes in the unified model.
An end-to-end ESS model was introduced in \cite{Lee17TacotronEmotion}, which focused on modifying Tacotron to get better alignment.
This paper instead studies how to build an end-to-end ESS model with limited emotion labels using GSTs.

\begin{figure}[t]
\centering
\includegraphics[width=0.7\linewidth]{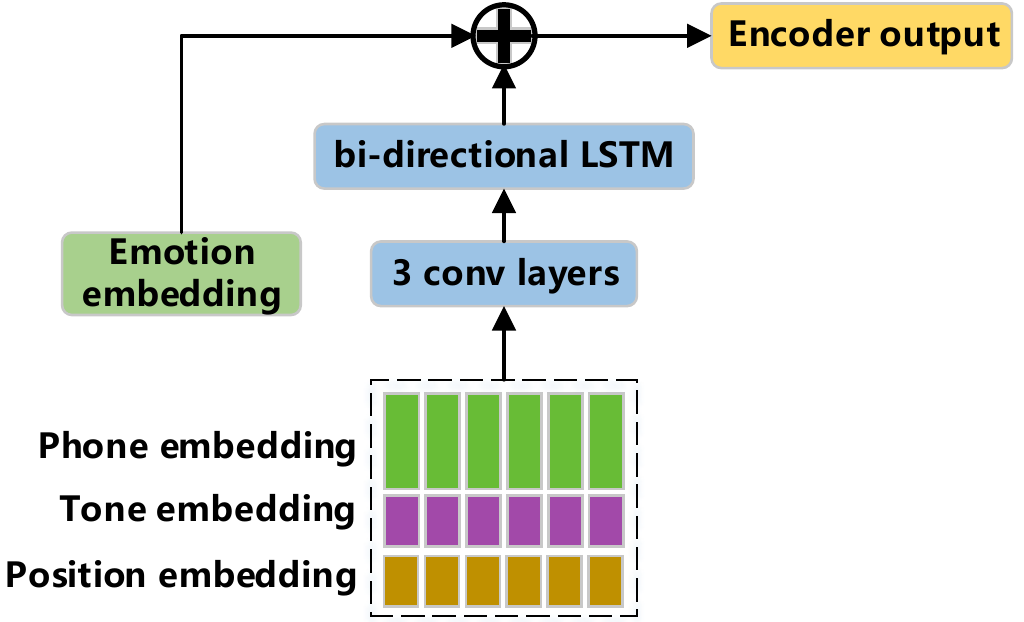} %
\caption{The architecture of the encoder using in our baseline Tacotron model.}
\label{fig:Baseline Encoder}
\end{figure}

\section{Methods}
\subsection{Tacotron-based acoustic model for ESS}\label{baseline}
This paper builds end-to-end ESS models based on Tacotron, and refers to an open source implementation\footnote{https://github.com/Rayhane-mamah/Tacotron-2.} of Tacotron 2.
Since our emotional speech corpus is recorded in Chinese, some modifications are made to the encoder module of Tacotron, as shown in Fig. \ref{fig:Baseline Encoder}. 
Instead of using Chinese character sequences directly, phoneme and tone sequences given by text analysis are adopted as input.
Here, phonemes are defined as the initials and finals of Chinese, and the tone of each phoneme is defined as the tone of the syllable in which current phoneme is located.
Besides, in order to get better break and rhythm performance, a feature which indicates the position of each phoneme in current sentence is added.
Therefore, for each phoneme, its representation is made up by concatenating a phone embedding, a tone embedding and a position embedding.

In order to build our Tacotron-based baseline models for ESS, we implement two approaches of emotional modeling, i.e., emotion-dependent (ED) modeling which trains a model for each emotion separately and emotion-independent (EI) modeling that trains a unified model for all emotions. In the EI model, an emotion embedding vector is added to the output of text encoder as shown in Fig. \ref{fig:Baseline Encoder}.

\begin{figure}[t]
\centering
\includegraphics[width=\linewidth]{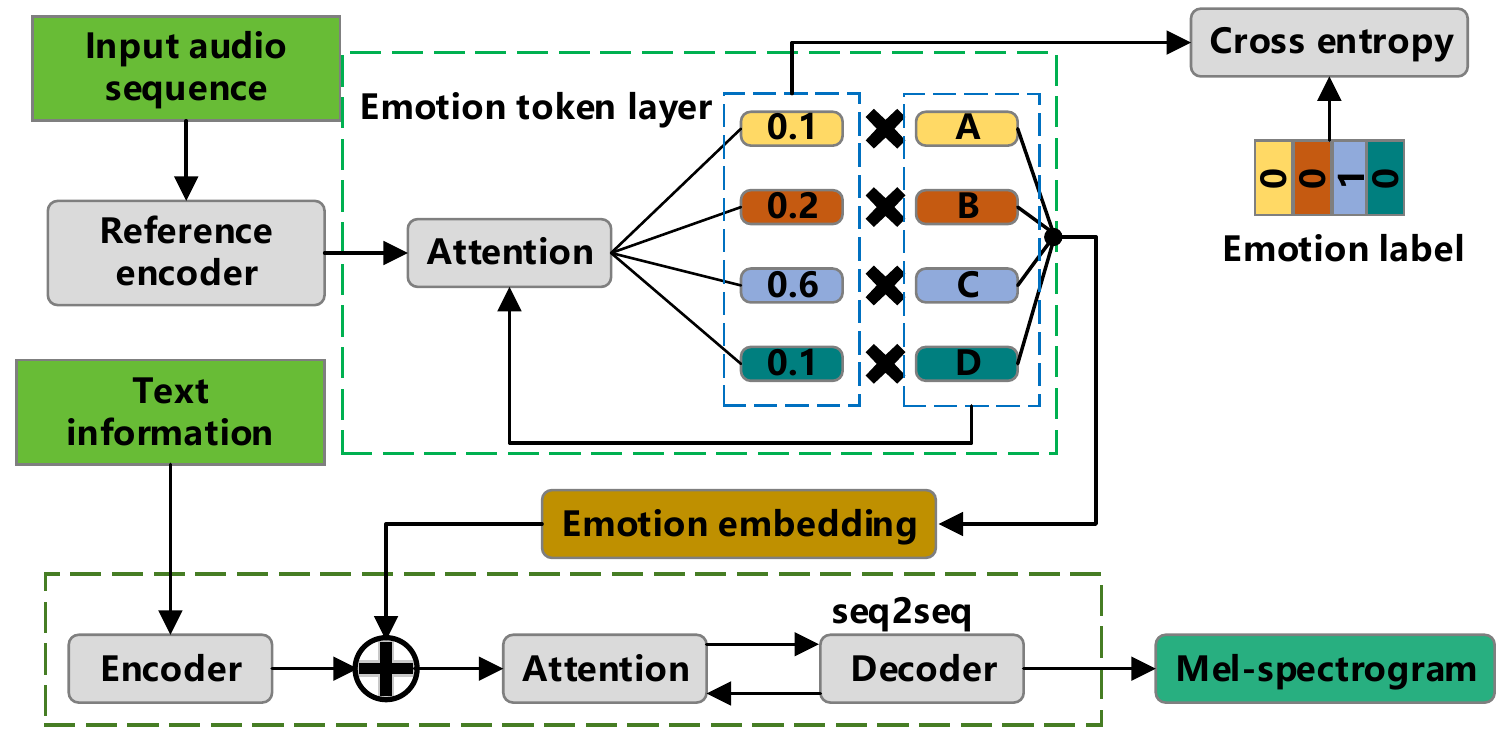} %
\caption{The model architecture of GST-Tacotron with semi-supervised training.}
\label{fig:main_struct}
\end{figure}

\subsection{GST-Tacotron with semi-supervised training for ESS}\label{proposed model}
This paper proposes a semi-supervised training method for Tacotron-based ESS.
This method adopts almost the same model structure as GST-Tacotron, including an emotion token layer and a Chinese Tacotron sequence-to-sequence module, as shown in Fig. \ref{fig:main_struct}.
The emotion token layer converts reference audio sequence into a fixed-length emotion embedding by two steps.
First, a reference encoder is utilized to compress the input audio sequence into a fixed-length vector.
Then, the vector is passed to the emotion token layer as the query vector for attention and to calculate a set of weights which measure its similarity to each token.
The weighted sum of tokens forms the emotion embedding which is added to the encoder output for every time step as shown in Fig. \ref{fig:Baseline Encoder}.

As introduced above, conventional GST-Tacotron models lack interpretability of the learnt tokens.
In order to guarantee that each learned token corresponds to one emotion in our method, several modifications are made to the conventional GST-Tacotron model.
First, single-head attention is used instead of multi-head attention in the emotion token layer to achieve a single-vector representation of token weights.
Second, the number of tokens is set to be consistent with the number of emotion categories in the corpus.
Finally, one-hot emotion labels are introduced into the emotion token layer as the target of token weights.
A cross entropy loss between the emotion labels and the token weights is added into the loss function of the GST-Tacotron model as follows,
\begin{equation}\label{eq1}
loss = \sum_{s}L(\textbf{c},\hat{\textbf{c}}) + \sum_{s{'}}CE(\textbf{e},\hat{\textbf{e}}),
\end{equation}
where $\textbf{c}$ and $\hat{\textbf{c}}$  denote the targets and predictions of an utterance in the training set,
$\textbf{e}$ and $\hat{\textbf{e}}$ represent the emotion label vector and the token weight vector of a training utterance respectively,
$L(\cdot)$ and $CE(\cdot)$ are Tacotron loss and cross entropy loss functions, $s$ denotes the complete training set and $s{'}$ denotes the part of training utterances with emotion labels.

At the synthesis stage, the emotion embedding is first determined by selecting the emotion token corresponding to the expected emotion.
Then, the embedding vector is added to the output of text encoder at each time step for following attention-based decoding.
For both the baseline method describe in Section \ref{baseline} and our proposed method,
a WaveNet vocoder  \cite{weiping18clarinet} conditioned on mel-spectrogram is trained to reconstruct waveform samples from the output of Tacotron.

\section{Experiments}
\subsection{Experimental conditions}\label{corpus}
A high-quality emotional speech corpus recorded in a professional studio was used in our experiments.
The 3057 Chinese sentences of  four emotions (neutral, happy, sad and angry) were uttered by a female speaker.
The recordings were about 4.48 hours with 16kHz sampling rate and 16bits quantization.
The details of the corpus are shown in Table \ref{corpus_table}.
\begin{table}[t]
\begin{center}

\caption{Details of the corpus used in our experiments. }
\begin{tabu}{c|cc|cc}
    \tabucline[0.75pt]{-}
    \multirow{2}{*}{emotion} & \multicolumn{2}{c|}{train}  & \multicolumn{2}{c}{test} \\
                      \cline{2-5}
                      & duration (min.)  & \# of utt.  & duration (min.)  & \# of utt. \\
                      \cline{1-5}
    neutral           & 84.68          & 759      & 3.20           & 30      \\
    happy             & 62.86          & 926      & 2.02           & 30      \\
    sad               & 87.33          & 882      & 3.02           & 30      \\
    angry             & 23.71          & 370      & 2.07          & 30      \\
     \cline{1-5}
    total             & 258.57          & 2937     & 10.30          & 120     \\
    \tabucline[0.75pt]{-}
\end{tabu}
\label{corpus_table}

\end{center}
\end{table}

Based on the speech corpus above, four types of models were constructed for comparison.
\begin{itemize}
  \item \textbf{ED}  Four emotion-dependent Tacotron models using all emotion labels as introduced in Section \ref{baseline}.
  \item \textbf{EI}  An emotion-independent Tacotron model using emotion embeddings and all emotion labels as introduced in Section \ref{baseline}.
  This model served as the upper-bound model for evaluating our proposed semi-supervised training method.
  \item \textbf{Semi-EI}  Same as \textbf{EI}, but only using the emotion labels of a part of training utterances.
  For the utterances without emotion labels, a zero emotion embedding vector is used.
  This model served as the baseline model for evaluating our proposed semi-supervised training method.
  \item \textbf{Semi-GST}  A GST-Tacotron model using the semi-supervised training method introduced in Section \ref{proposed model}.
\end{itemize}

The output acoustic features of all models were 80-dimensional mel-spectrogram.
The dimensions of phone embedding, tone embedding and position embedding  were 384, 64 and 64 respectively.
The settings of Tacotron models and the emotion token layer in Semi-GST followed previous studies \cite{Shen18tacotron2,Wang18GSTs}.
The WaveNet neural vocoder had 24 dilated casual convolution layers which were divided into 4 convolution blocks. Each block contained 6 layers and their dilation coefficients were \{1,2,4,8,16,32\}. The output distribution for each waveform sample was single Gaussian.
When building the ED models, it was difficult to get Tacotron models which can generate normal alignment between phoneme and acoustic feature sequences since the training data of a single emotion was limited.
Therefore, a pre-training and adaptation strategy was utilized. First, a unified Tacotron model was estimated with 1000 epochs using all training data without emotion labels.
Then, the data of each emotion was utilized to fine-tune the model with 100 epochs to get corresponding ED model.
All the EI, Semi-EI and Semi-GST models were trained with 1000 epochs.

\begin{table}[t]
\begin{center}
\caption{Objective evaluation results of ED and EI models.}
\begin{tabular}{c c c c c c c}
\toprule
    \textbf{Model} & \textbf{Emotion} &\textbf{MCD(dB)} & \textbf{F$_{0}$RMSE(Hz)} &  \textbf{V/UV(\%)} & \textbf{FFE(\%)}\\
    \midrule
        \multirow{4}{*}{ED} & neutral & 2.75 & 57.54 & 7.94   & 20.00  \\
                            & happy   & 2.65 & 73.87 & 9.06   & 25.14  \\
                            & sad     & 2.61 & 53.99 & 6.95   & 17.24  \\
                            & angry   & 2.69 & 77.88 & 10.90  & 28.68  \\
         \midrule
        \multirow{4}{*}{EI} & neutral & 2.66 & 58.19 & 7.20   & 20.19 \\
                            & happy   & 2.63 & 74.59 & 8.53   & 26.30 \\
                            & sad     & 2.54 & 51.47 & 6.82   & 15.48 \\
                            & angry   & 2.61 & 76.94 & 10.34  & 27.31 \\
\bottomrule
\end{tabular}
\label{EI_ED_obj_eval}
\end{center}
\end{table}

\subsection{Comparison between ED and EI models}
Objective evaluations were conducted to compare the performance of ED and EI models, which both followed the supervised training framework.
The evaluation metrics were the average distortions of acoustic feature prediction,
including mel-cepstrum distortions (MCD), root mean square error of F$_{0}$ (F$_{0}$RMSE), voiced/unvoiced decision error (V/UV) and F$_{0}$ frame error (FFE) \cite{weichu2009FFE}, on the test set of each emotion. The FASTDTW \cite{salvador2004fastdtw} algorithm was adopted to align the predicted acoustic feature sequences towards the natural ones.
Then, the distortions between them were calculated frame-by-frame.

The results are shown in Table \ref{EI_ED_obj_eval}.
From the table, we can see that in both  ED and EI models,  ``neutral" and ``sad" emotions obtained better results than ``happy" and ``angry" emotions on almost all metrics.
It can be explained from two aspects. First, the later two emotions had higher and more variational F0 contours which were more difficult for modeling.
Second, the ``angry" emotion had much less training data as shown in Table \ref{corpus_table}.
Comparing  ED and EI models, it can be observed that the EI model achieved slightly better results than ED for all emotions.
This can be attributed to the advantage of unified modeling and data sharing.

\subsection{Objective evaluation on our proposed method}
First, we conducted an emotion recognition experiment to verify the interpretability of the tokens learned by the Semi-GST model.
Four Semi-GST models were trained which utilized 2\%, 5\%, 10\% and 20\% emotion labels in the training set respectively.
In this experiment, we fed the 120 test utterances (30 per emotion) in the test set into the emotion token layer and got 120 token weight vectors.
For each vector, the index of its maximum value was treated as the emotion recognition result of this utterance.
Table \ref{confusion_matrix} shows the confusion matrices of emotion recognition.
The $\bar{w}$ value in Table \ref{confusion_matrix} denotes the weight of the token corresponding to the true emotion and averaged across the 30 test utterances.

We expect that all $\bar{w}$ values are close to 1 which indicates a one-to-one correspondence between the learnt tokens and the emotion categories.
\begin{table}[t]
\caption{Confusion matrices of emotion recognition using the Semi-GST models trained with different proportions of emotion labels.
Here, $\bar{w}$ stands for the average token weight of the true emotion.}
\centering
\subtable[2\% emotion labels]{
       \begin{tabu}{c|ccccc}
       \tabucline[0.75pt]{-}
            \multirow{2}{*}{\textbf{true label}}  & \multicolumn{5}{c}{\textbf{recognized label}}       \\ \cline{2-6}
            & neutral & happy & sad & angry & $\bar{w}$ \\
            \cline{1-6}
            neutral                     & \textbf{26}      & 0      & 4   &   0    & 0.8554  \\
            happy                       & 2       & \textbf{21}    & 7   &    0   & 0.7387  \\
            sad                         & 4       & 5     & \textbf{21}  &    0   & 0.6553  \\
            angry                       & 2       & 8     & \textbf{20}  &     0  & 0.0336  \\
            \tabucline[0.75pt]{-}
            \end{tabu}
}
\qquad
\subtable[5\% emotion labels]{
       \begin{tabu}{c|ccccc}
       \tabucline[0.75pt]{-}
            \multirow{2}{*}{\textbf{true label}}  & \multicolumn{5}{c}{\textbf{recognized label}}       \\ \cline{2-6}
            & neutral & happy & sad & angry & $\bar{w}$ \\
            \cline{1-6}
            neutral                     & \textbf{29}      &  1     &   0  &   0      & 0.9598  \\
            happy                       &  0       & \textbf{30}     &  0   &   0      & 0.9927  \\
            sad                         &   0      & 1      & \textbf{29}  &    0     & 0.9507  \\
            angry                       &   0      &   0     &  0   &     \textbf{30} & 1.0000  \\
            \tabucline[0.75pt]{-}
            \end{tabu}
}
\qquad
\subtable[10\% emotion labels]{
       \begin{tabu}{c|ccccc}
       \tabucline[0.75pt]{-}
            \multirow{2}{*}{\textbf{true label}}  & \multicolumn{5}{c}{\textbf{recognized label}}       \\ \cline{2-6}
            & neutral & happy & sad & angry & $\bar{w}$ \\
            \cline{1-6}
            neutral                     & \textbf{30}      &   0     &   0  &  0       & 0.9890  \\
            happy                       &   0      & \textbf{30}     &   0  &  0       & 0.9999  \\
            sad                         &   0      & 1      & \textbf{29}  &    0     & 0.9667  \\
            angry                       &   0      &  0      &  0   &      \textbf{30} & 1.0000  \\
            \tabucline[0.75pt]{-}
            \end{tabu}
}
\qquad
\subtable[20\% emotion labels]{
       \begin{tabu}{c|ccccc}
       \tabucline[0.75pt]{-}
            \multirow{2}{*}{\textbf{true label}}  & \multicolumn{5}{c}{\textbf{recognized label}}       \\ \cline{2-6}
            & neutral & happy & sad & angry & $\bar{w}$ \\
            \cline{1-6}
            neutral                     & \textbf{30}      &  0      &  0   &   0      & 1.0000  \\
            happy                       &  0       & \textbf{30}     &  0   &   0      & 1.0000  \\
            sad                         &  0       &  0      & \textbf{30}  &   0      & 0.9993  \\
            angry                       &  0      &   0     &   0  &      \textbf{30} & 1.0000  \\
            \tabucline[0.75pt]{-}
            \end{tabu}
}
\label{confusion_matrix}
\end{table}

From Table \ref{confusion_matrix}, we can see that there were plenty of recognition errors when only 2\% emotion labels were utilized.
When increasing the proportion of available emotion labels to 5\%, the confusion matrix appeared in a clear diagonal form and
the $\bar{w}$ values became very close to 1.
Further, there were no classification errors when using 20\% emotion labels.
These results reflect that the Semi-GST model can achieve the emotion-related interpretability of tokens with only 5\% emotion labels.

\begin{table}[t]
\begin{center}
\begin{threeparttable}
\caption{Objective evaluation results of EI, Semi-EI and Semi-GST models.}
\begin{tabular}{ c c c c c}
\toprule
    \textbf{Model} & \textbf{MCD(dB)} & \textbf{F$_{0}$RMSE(Hz)} &  \textbf{V/UV(\%)} & \textbf{FFE(\%)}\\
    \midrule
             EI             & {2.61} & {63.77} & {7.98}   & {21.66}  \\
          Semi-EI       & 2.71 & 68.95 & 8.97   & 24.93  \\ 		
          Semi-GST       & 2.64 & 64.04 & 8.26   & 21.81  \\

\bottomrule
\end{tabular}
\label{model_obj_compare}
\end{threeparttable}
\end{center}
\end{table}
\begin{table}[t]
\begin{center}
\begin{threeparttable}
\caption{Average preference scores(\%) among EI, Semi-EI and Semi-GST models on naturalness of synthetic speech, where N/P stands for ``no preference" and $p$ denotes the $p$-value of a $t$-test between two models.}
\begin{tabular}{ c c c c c}
\toprule
     \textbf{Semi-GST} & \textbf{EI} &  \textbf{Semi-EI} & \textbf{N/P} & $\bm{p}$\\
    \midrule
          66.25 & - & 18.33   & 15.42 &   $2.30\times10^{-34}$ \\ 		
          35.00 & 38.96 & -   & 26.04 &   0.31 \\

\bottomrule
\end{tabular}
\label{ABX}
\end{threeparttable}
\end{center}
\end{table}

\begin{table}[t]
\caption{The percentages of perceived emotions for speech synthesized using different models.}
\centering
\subtable[EI]{
       \begin{tabu}{c|ccccc}
       \tabucline[0.75pt]{-}
            \multirow{2}{*}{\textbf{true label}}  & \multicolumn{5}{c}{\textbf{perceived label}}       \\ \cline{2-6}
            & neutral & happy & sad & angry & other \\
            \cline{1-6}
            neutral                     & \textbf{91.67}       & 5.00       &    0     &  0.83      & 2.50  \\
            happy                       & 12.50       & \textbf{85.00}      &    0     &   0         & 2.50  \\
            sad                         &   0          &    0        & \textbf{100.00}  &   0         &  0   \\
            angry                       & 5.83        & 2.50       &   0      &   \textbf{89.17}    & 2.50  \\
            \tabucline[0.75pt]{-}
            \end{tabu}
}
\qquad
\subtable[Semi-EI]{
       \begin{tabu}{c|ccccc}
       \tabucline[0.75pt]{-}
            \multirow{2}{*}{\textbf{true label}}  & \multicolumn{5}{c}{\textbf{perceived label}}       \\ \cline{2-6}
            & neutral & happy & sad & angry & other \\
            \cline{1-6}
            neutral                     & \textbf{87.50}       & 9.17      &  0.83      &  0.83       & 1.67  \\
            happy                       & 5.83        & \textbf{91.67}     &  0          &    0         & 2.50  \\
            sad                         & 1.67        & 0          &  \textbf{96.67\ \ }     &   0.83      &  0.83     \\
            angry                       & 9.17        & 15.00     &  0.83      &   \textbf{69.17}     & 5.83  \\
            \tabucline[0.75pt]{-}
            \end{tabu}
}
\qquad
\subtable[Semi-GST]{
       \begin{tabu}{c|ccccc}
       \tabucline[0.75pt]{-}
            \multirow{2}{*}{\textbf{true label}}  & \multicolumn{5}{c}{\textbf{perceived label}}       \\ \cline{2-6}
            & neutral & happy & sad & angry & other \\
            \cline{1-6}
            neutral                     & \textbf{92.50}       & 2.50       & 1.67        &  0.83       & 2.50  \\
            happy                       & 4.17        & \textbf{94.16}      & 0            &  0           & 1.67  \\
            sad                         & 0            & 0           & \textbf{100.00}      & 0           & 0      \\
            angry                       & 8.33        & 5.83       &   0.83      &   \textbf{80.83}     & 4.17  \\
            \tabucline[0.75pt]{-}
            \end{tabu}
}
\label{classification_rate}
\end{table}
Then, an experiment was conducted to compare the performance of EI, Semi-EI and Semi-GST models. 
In this experiment, 5\% emotion labels were used for both Semi-EI and Semi-GST models.
The metrics were the same as the ones used in Table \ref{EI_ED_obj_eval} and the results averaged across  all emotions are shown in Table \ref{model_obj_compare}.
We can see that Semi-GST performed better than Semi-EI on all metrics. The performance of the Semi-GST model was very close to that of the EI model although it only utilized 5\% emotion labels.

\subsection{Subjective evaluation on our proposed method}
To subjectively evaluate the performance of our proposed method on the naturalness and emotion expressiveness of synthetic speech\footnote{The audio samples can be found at
\url{http://home.ustc.edu.cn/~wpf0610/APSIPA2019.html}},
two groups of ABX preference tests and three emotion classification tests were conducted by 12 subjects. In both subjective tests, ten sentences were selected randomly for each emotion from the synthesized test set.

In ABX preference tests, the subjects were asked to give one of the three choices, (1) A is more natural, (2) no preference, (3) B is more natural, for each pair of synthetic speech played in random order.
In addition to the average preference scores, the $p$-value of $t$-test was used to measure the significance of the difference between two models in each test.
The results are shown in Table \ref{ABX}.
We can see that Semi-GST significantly outperformed Semi-EI ($p < 0.01$) and there was no significant difference between Semi-GST and EI ($p =0.31$).

In emotion classification tests, the synthetic utterances of all emotions were played in random order.
Each subject was asked to choose the emotion they perceived for each utterance. They were also allowed to chose the ``other" option if they didn't think the utterance belonged to any of the the four emotions.
The results are shown in Table \ref{classification_rate}.
First, we can see that all these three models achieved the best accuracy for the ``sad" emotion and the worst accuracy for the ``angry" emotion.
This is consistent with the objective evaluation results shown in Table \ref{EI_ED_obj_eval}.
Further, both EI and Semi-GST models performed better than the Semi-EI model.
The average classification accuracies of the three models were  91.46\% (EI), 86.25\% (Semi-EI), and 91.88\%(Semi-GST) respectively,
which shows that the Semi-GST model can express emotions as effectively as the EI model.

\section{Conclusions}
This paper have presented an end-to-end emotional speech synthesis method using style tokens and semi-supervised training.
The model is implemented based on the GST-Tacotron framework.
A cross-entropy loss function was introduced to achieve the emotion-related interpretability of style tokens and the semi-supervised training using a small portion of emotion labels in the training data.
Experimental results confirmed the effectiveness of our proposed method.
Objective and subjective evaluation results demonstrated that our model outperformed the conventional Tacotron model for emotional speech synthesis when only 5\% of training
data has emotion labels. The naturalness and emotion expressiveness of our proposed method using 5\% emotion labels were close to the Tacotron model using all emotion labels.
To apply our proposed method to spontaneous emotional and expressive speech will be a task of our future work.

\section*{Acknowledgment}
This work was supported by the National Key R\&D Program of China (Grant No. 2017YFB1002202), the National Nature Science Foundation of China (Grant No. 61871358 and U1613211).

\bibliographystyle{IEEEtran}

\bibliography{mybib}

\begin{thebibliography}{10}
\providecommand{\url}[1]{#1}
\csname url@samestyle\endcsname
\providecommand{\newblock}{\relax}
\providecommand{\bibinfo}[2]{#2}
\providecommand{\BIBentrySTDinterwordspacing}{\spaceskip=0pt\relax}
\providecommand{\BIBentryALTinterwordstretchfactor}{4}
\providecommand{\BIBentryALTinterwordspacing}{\spaceskip=\fontdimen2\font plus
\BIBentryALTinterwordstretchfactor\fontdimen3\font minus
  \fontdimen4\font\relax}
\providecommand{\BIBforeignlanguage}[2]{{%
\expandafter\ifx\csname l@#1\endcsname\relax
\typeout{** WARNING: IEEEtran.bst: No hyphenation pattern has been}%
\typeout{** loaded for the language `#1'. Using the pattern for}%
\typeout{** the default language instead.}%
\else
\language=\csname l@#1\endcsname
\fi
#2}}
\providecommand{\BIBdecl}{\relax}
\BIBdecl

\bibitem{ZeSS13}
H.~Zen, A.~W. Senior, and M.~Schuster, ``Statistical parametric speech
  synthesis using deep neural networks,'' in \emph{{IEEE} International
  Conference on Acoustics, Speech and Signal Processing, {ICASSP} 2013,
  Vancouver, BC, Canada, May 26-31, 2013}, 2013, pp. 7962--7966.

\bibitem{44312}
H.~Zen, ``Statistical parametric speech synthesis: from hmm to lstm-rnn,''
  2015, lecture given at RTTH Summer School on Speech Technology, Barcelona,
  Spain.

\bibitem{YF2014SSbiLSTM}
Y.~Fan, Y.~Qian, F.~Xie, and F.~K. Soong, ``{TTS} synthesis with bidirectional
  {LSTM} based recurrent neural networks,'' in \emph{{INTERSPEECH} 2014, 15th
  Annual Conference of the International Speech Communication Association,
  Singapore, September 14-18, 2014}, 2014, pp. 1964--1968.

\bibitem{vande2016wavenet}
A.~van~den Oord, S.~Dieleman, H.~Zen, K.~Simonyan, and O.~V. et~al., ``Wavenet:
  {A} generative model for raw audio,'' \emph{CoRR}, vol. abs/1609.03499, 2016.

\bibitem{Wang17tacotron}
Y.~Wang, R.~J. Skerry{-}Ryan, D.~Stanton, Y.~Wu, and R.~J.~W. et~al.,
  ``Tacotron: Towards end-to-end speech synthesis,'' in \emph{Interspeech 2017,
  18th Annual Conference of the International Speech Communication Association,
  Stockholm, Sweden, August 20-24, 2017}, 2017, pp. 4006--4010.

\bibitem{Shen18tacotron2}
J.~Shen, R.~Pang, R.~J. Weiss, M.~Schuster, and N.~J. et~al., ``Natural {TTS}
  synthesis by conditioning wavenet on {MEL} spectrogram predictions,'' in
  \emph{2018 {IEEE} International Conference on Acoustics, Speech and Signal
  Processing, {ICASSP} 2018, Calgary, AB, Canada, April 15-20, 2018}, 2018, pp.
  4779--4783.

\bibitem{HoferRC05interspeech}
G.~Hofer, K.~Richmond, and R.~A.~J. Clark, ``Informed blending of databases for
  emotional speech synthesis,'' in \emph{{INTERSPEECH} 2005 - Eurospeech, 9th
  European Conference on Speech Communication and Technology, Lisbon, Portugal,
  September 4-8, 2005}, 2005, pp. 501--504.

\bibitem{Cahn90thegeneration}
J.~Cahn, ``The generation of affect in synthesized speech,'' \emph{Journal of
  the American Voice I/O Society}, vol.~8, pp. 1--19, 1990.

\bibitem{MurrayA95FSE}
I.~R. Murray and J.~L. Arnott, ``Implementation and testing of a system for
  producing emotion-by-rule in synthetic speech,'' \emph{Speech Communication},
  vol.~16, no.~4, pp. 369--390, 1995.

\bibitem{Schroder01ESSreview}
M.~Schr{\"{o}}der, ``Emotional speech synthesis: a review,'' in
  \emph{{EUROSPEECH} 2001 Scandinavia, 7th European Conference on Speech
  Communication and Technology, 2nd {INTERSPEECH} Event, Aalborg, Denmark,
  September 3-7, 2001}, 2001, pp. 561--564.

\bibitem{YamagishiOMK05}
J.~Yamagishi, K.~Onishi, T.~Masuko, and T.~Kobayashi, ``Acoustic modeling of
  speaking styles and emotional expressions in hmm-based speech synthesis,''
  \emph{{IEICE} Transactions}, vol. 88-D, no.~3, pp. 502--509, 2005.

\bibitem{NoseYMK07}
T.~Nose, J.~Yamagishi, T.~Masuko, and T.~Kobayashi, ``A style control technique
  for hmm-based expressive speech synthesis,'' \emph{{IEICE} Transactions},
  vol. 90-D, no.~9, pp. 1406--1413, 2007.

\bibitem{AnLD17LSTM}
S.~An, Z.~Ling, and L.~Dai, ``Emotional statistical parametric speech synthesis
  using lstm-rnns,'' in \emph{2017 Asia-Pacific Signal and Information
  Processing Association Annual Summit and Conference, {APSIPA} {ASC} 2017,
  Kuala Lumpur, Malaysia, December 12-15, 2017}, 2017, pp. 1613--1616.

\bibitem{Lee17TacotronEmotion}
Y.~Lee, A.~Rabiee, and S.~Lee, ``Emotional end-to-end neural speech
  synthesizer,'' \emph{CoRR}, vol. abs/1711.05447, 2017.

\bibitem{LT18DNNEmotion}
J.~Lorenzo{-}Trueba, G.~E. Henter, S.~Takaki, J.~Yamagishi, and Y.~M. et~al.,
  ``Investigating different representations for modeling and controlling
  multiple emotions in dnn-based speech synthesis,'' \emph{Speech
  Communication}, vol.~99, pp. 135--143, 2018.

\bibitem{Skerry18tacotron}
R.~J. Skerry{-}Ryan, E.~Battenberg, Y.~Xiao, Y.~Wang, and D.~S. et~al.,
  ``Towards end-to-end prosody transfer for expressive speech synthesis with
  tacotron,'' in \emph{Proceedings of the 35th International Conference on
  Machine Learning, {ICML} 2018, Stockholmsm{\"{a}}ssan, Stockholm, Sweden,
  July 10-15, 2018}, 2018, pp. 4700--4709.

\bibitem{Wang18GSTs}
Y.~Wang, D.~Stanton, Y.~Zhang, R.~J. Skerry{-}Ryan, and E.~B. et~al., ``Style
  tokens: Unsupervised style modeling, control and transfer in end-to-end
  speech synthesis,'' in \emph{Proceedings of the 35th International Conference
  on Machine Learning, {ICML} 2018, Stockholmsm{\"{a}}ssan, Stockholm, Sweden,
  July 10-15, 2018}, 2018, pp. 5167--5176.

\bibitem{AkuzawaIM18VAE}
K.~Akuzawa, Y.~Iwasawa, and Y.~Matsuo, ``Expressive speech synthesis via
  modeling expressions with variational autoencoder,'' in \emph{Interspeech
  2018, 19th Annual Conference of the International Speech Communication
  Association, Hyderabad, India, 2-6 September 2018.}, 2018, pp. 3067--3071.

\bibitem{weiping18clarinet}
\BIBentryALTinterwordspacing
W.~Ping, K.~Peng, and J.~Chen, ``Clarinet: Parallel wave generation in
  end-to-end text-to-speech,'' \emph{CoRR}, vol. abs/1807.07281, 2018.
  [Online]. Available: \url{http://arxiv.org/abs/1807.07281}
\BIBentrySTDinterwordspacing

\bibitem{weichu2009FFE}
{Wei Chu} and A.~{Alwan}, ``Reducing {F}0 frame error of {F}0 tracking
  algorithms under noisy conditions with an unvoiced/voiced classification
  frontend,'' in \emph{2009 IEEE International Conference on Acoustics, Speech
  and Signal Processing}, 2009, pp. 3969--3972.

\bibitem{salvador2004fastdtw}
S.~Salvador and P.~Chan, ``Fastdtw: Toward accurate dynamic time warping in
  linear time and space,'' in \emph{KDD workshop on mining temporal and
  sequential data}.\hskip 1em plus 0.5em minus 0.4em\relax Citeseer, 2004.

\end{thebibliography}
\end{document}